\documentclass[conference]{IEEEtran}

\usepackage{amsmath}

\DeclareMathOperator*{\argmin}{arg\,min}

\DeclareMathOperator*{\tr}{tr}
\DeclareMathOperator*{\E}{E}
\newcommand{\T}{\operatorname{\mathrm{T}}}

\newcommand{\He}{\text{H}}
\newcommand{\diag}{\text{diag}}

\newcommand{\sign}{\text{sign}}

\newcommand{\vect}[1]{\mathbf{#1}}
\newcommand{\matt}[1]{\mathbf{#1}}

\usepackage{tikz}
\usetikzlibrary{arrows,snakes,shapes,positioning,arrows,decorations.markings,calc}
\usetikzlibrary{dsp,chains}
\usepackage{psfrag}

\usepackage{pgfplots}

\usepackage{comment}
\usepackage{algorithm}
\usepackage{algorithmic}

\usepackage{authblk}
\begin{document}

\title{Precoding under Instantaneous \\
Per-Antenna Peak Power Constraint}

\author[1]{Hela~Jedda}
\author[2]{Amine~Mezghani}
\author[2]{A.~Lee Swindlehurst}
\author[1,3]{Josef~A.~Nossek }
\affil[1]{Technical University of Munich, 80290 Munich, Germany}
\affil[2]{University of California, Irvine, Irvine, CA 92697, USA}
\affil[3]{Federal University of Cear\'a, Fortaleza, Brazil}
\affil[ ]{Email: hela.jedda@tum.de, amezghan@uci.edu, swindle@uci.edu, josef.a.nossek@tum.de}

\maketitle

\tikzset{DSP lines/.style={help lines,very thick,color=black}}
\tikzset{line_arrow/.style={help lines,very thick,color=black,->,-angle 90}}
\tikzset{filter/.style={rectangle,inner sep=0pt,minimum height=0.8cm,minimum width=1cm,draw=black,very thick}}
\tikzset{delay/.style={rectangle,inner sep=0pt,minimum size=1cm,draw=black,very thick}}
\tikzset{downsampling/.style={rectangle,inner sep=0pt,minimum height=0.8cm,minimum width=0.7cm,draw=black,very thick}}
\tikzset{upsampling/.style={rectangle,inner sep=0pt,minimum height=0.8cm,minimum width=0.7cm,draw=black,very thick}}
\tikzset{empty_node/.style={inner sep=0pt,minimum size=0cm}}
\tikzset{connection/.style={circle,draw=black,fill=black,inner sep=0pt,minimum size=2mm}}
\tikzset{coefficient/.style={isosceles triangle,draw=black,very thick,inner sep=0pt,minimum size=.7cm}}
\tikzset{source/.style={semicircle,minimum size=.5cm,draw=black,very thick,shape border rotate=270}}
\tikzset{adder/.style={circle,minimum size=.25cm,inner sep=0pt,draw=black,very thick}}
\tikzset{multiplier/.style={circle,minimum size=.25cm,inner sep=0pt,draw=black,very thick}}
\tikzset{double_arrow/.style={double distance=5pt,thick,shorten >= 6pt,decoration={markings,mark=at position 1 with {\arrow[scale=.6,>=angle 90]{>}}},postaction={decorate}}}
\begin{abstract}
We consider a multi-user (MU) multiple-input-single-output (MISO) downlink system with $M$ single-antenna users and $N$ transmit antennas with a nonlinear power amplifier (PA) at each antenna. Instead of emitting constant envelope (CE) signals from the antennas to have highly power efficient PAs, we relax the CE constraint and allow the transmit signals to have instantaneous power less than or equal to the available power at each PA. The PA power efficiency decreases but simulation results show that the same performance in terms of bit-error-ratio (BER) can be achieved with less transmitted power and less PA power consumption. We propose a linear and a nonlinear precoder design to mitigate the multi-user interference (MUI) under the constraint of a maximal instantaneous per-antenna peak power.
\end{abstract}


%
\IEEEpeerreviewmaketitle

\section{Introduction}
Power efficiency is a big concern in future communication systems. As the power amplifier (PA) typically accounts for more than half of the power consumption in a base station (BS) \cite{Blume2010, Chen2010}, it is desirable in terms of high power efficiency to run the PA in the saturation region. In this way the power consumed by the PA is totally radiated and no power is lost as heat. Operation in the saturation region, however, implies high distortions and nonlinearities that are introduced to the signals. One way to avoid these distortions is to design the signals to be constant envelope (CE) at the PA input \cite{MohammedLarsson2012, Larsson2013, MohammedLarsson2013, MollénLarsson2015, prabhu2015, AmadoriMasouros2016}. However,  these works guarantee the CE property only at the discrete symbol time, which might moderate the actual PA efficiency improvement at the end. This adds also to the fact that, for certain amplifier implementation like push-pull (class B) PA, the efficiency increases only subproportionally to the peak-to-average power ratio reduction at the input. Moreover, emitting signals with the maximal available power does not necessarily lead to better communications performance. More important than running the PA efficiently is to use the transmit power and the PA power efficiently to achieve better system performance.
When using CE signals, the power allocation at the transmit antennas is constant and each antenna transmits with the maximal available power, independent of the channel. Is it not better to allocate less power or no power at some antennas if they lead to only destructive interference at the receiver or to contaminate the environment with unuseful power?  In fact the use of the linear region of the PA might be inevitable due to spectral shaping requirements. The PA power efficiency decreases but how efficiently are the radiated power and the PA power used to achieve a certain BER?

In this work, we aim at answering these questions. To this end, and based on the above, we relax the CE constraint and let the instantaneous power at each antenna be less than or equal to the maximal available power for each antenna and consider the required system power needed to achieve a certain BER value. In this case, the PA operating region is not restricted to the saturation region but is extended to a linear region and a saturation region. In this context, the authors in \cite{Iofedov2015} considered the problem of precoder design based on maximum-ratio transmission with PA distortions in single-user OFDM systems.

We consider a downlink multi-user (MU) multiple-input-single-output (MISO) scenario. The PA is modeled using a clipping function for large input amplitudes. The goal is to design a precoder to mitigate the multi-user interference (MUI) as well as the PA nonlinearities.

This paper is organized as follows. In Section \ref{sec:sysmodel} we present the system model. Section \ref{sec:pamodel} describes the PA model that we consider throughout our work. The linear and the nonlinear optimization problems are presented in Sections \ref{sec:linearprecoder} and \ref{sec:nonlinearprecoder}, and algorithms are developed to solve them. In Sections \ref{sec:results} and \ref{sec:conclusion} we discuss the simulation results and summarize this work.

\textbf{Notation}: Bold letters indicate vectors and matrices, non-bold letters express scalars. The operators $(.)^{*}$, $(.)^{\rm T}$, $(.)^{\rm H}$ and $\E\left\lbrace \bullet \right\rbrace $ stand for complex conjugation, transposition, Hermitian transposition and expectation, respectively. The $n \times n$ identity (zeros) matrix is denoted by $\mathbf{I}_{n}$ ($\mathbf{0}_{n}$).  The vector $\vect{e}_{l}$ represents a zero vector with 1 in the $l$-th position. $\diag(\matt{A})$ denotes a diagonal matrix containing only the diagonal elements of $\matt A$.

\section{System Model}
\label{sec:sysmodel}

\begin{figure}[h!]
	\centering
	\resizebox{9cm}{!} {%
\begin{tikzpicture}

\node (in){};
\node[filter] (mapping) [right=of in] {$ \matt P$};
\node[filter] (quantizer1) [right=of mapping] {$g_{\text{PA}}(\bullet)$};
\node[filter] (channel) [right=of quantizer1]  {$\matt H$};
\node[adder] (noise) [right=of channel]   {$+$};
\node (n)[below=of noise][yshift=0.5cm] {$\boldsymbol{\eta}$};
\node[filter] (quantizer2) [right=of noise][xshift=-0.5cm]{$\matt F$};
\node (out) [right=of quantizer2]{};

\draw[DSP lines] [-stealth] (in.east) -- (mapping.west)node[pos=0.5,above]{$\vect s$}node[pos=0.5,below]{$M$} ;
\draw[DSP lines] [-stealth] (mapping.east) -- (quantizer1.west)  node[pos=0.5,above]{$\vect x$} node[pos=0.5,below]{$N$} ;
\draw[DSP lines] [-stealth] (quantizer1.east) -- (channel.west)node[pos=0.5,above]{$\vect x_{\text{PA}}$}node[pos=0.5,below]{$N$};
\draw[DSP lines] [-stealth] (channel.east) -- (noise.west)node[pos=0.5,below]{$M$};
\draw[DSP lines] [-stealth] (n) -- (noise.south);
\draw[DSP lines] [-stealth] (noise.east) -- (quantizer2.west)node[pos=0.5,above]{$\vect { r}$};
\draw[DSP lines] [-stealth] (quantizer2.east) -- (out) node[pos=0.5,above]{$\vect {\hat s}$}node[pos=0.5,below]{$M$}  ;

\end{tikzpicture}%
}
	\caption{System model for the MU-MISO case.}
	\label{fig:sysmodel}
\end{figure}
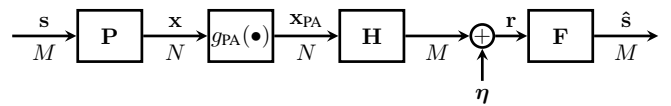

We consider a downlink MU-MISO system as depicted in Fig. \ref{fig:sysmodel}. The BS has $N$ antennas and serves $M$ users  each with a single antenna. The vector $\vect s$ of dimension $M$ contains the symbols for each user. The symbols are drawn from $B$-QAM constellation and are independent identically distributed (i.i.d.) with zero mean and covariance matrix $\matt C_{\vect s}= \sigma_s^2\matt{I}_M$. The input signal is processed by the precoder to get the $N$-dimensional signal vector $\vect x$ that goes into the PA. The considered PA model is presented in Section \ref{sec:pamodel}.

The received signal reads as $\vect{\hat s} = \matt F \left(\matt {H}g_{\text{PA}}(\vect x)+\boldsymbol{\eta}\right)$, where $\matt F$ is a diagonal real-valued matrix, $\matt {H}$ represents the channel matrix with i.i.d. Rayleigh-fading channel coefficients of unit variance and $\boldsymbol{\eta} \sim \mathcal{CN}(\vect{0}_M,\matt C_{\boldsymbol \eta}=\matt{I}_M)$ is the noise vector. 
The diagonal elements of $\matt F$ represent the gain that must be applied at the receivers in order to optimally decode the transmitted symbols.
The precoder $\matt P$ as well as the receive gains $\matt F$ have to be designed to mitigate the MUI and the PA distortions.
For the precoder design at the transmitter we assume $\matt F = f \matt I_M$ to ensure some degree of fairness between users. Once the designed precoder is applied, at the receive side each user can compute a new optimized $f_m, m=1,\cdots,M$, independently of the other users.

\section{PA Model}
\label{sec:pamodel}
The considered PA is a nonlinear device which clips the output signal if the input signal is larger than a certain saturation value. The PA nonlinearities cause interference and distortion, which create in-band and out-of-band spectral spread \cite{morgan2006generalized}. To quantify its power consumption $P_{\text{PA}}$, a model of the PA has to be first introduced and then a closed form expression for calculation of $P_{\text{PA}}$ can be given.
The dominant source of PA distortion is the amplitude distortion or AM-AM conversion. It describes the relation between the amplitudes of the PA's input and output signals \cite{ali2008behavioral}. The AM-AM distortion can incorporate most of the PA nonlinear effects \cite{liu2005practical}. The PA is modeled as a transformer based push-pull stage with the following approximate characteristic
\begin{align}
g_{\text{PA}}(x_n)= \begin{cases} x_n & \text{ if } \vert x_n \vert \leq \sqrt{\frac{P_{\text{tx}}}{N}}, \\
\sign(x_n) \sqrt{\frac{P_{\text{tx}}}{N}} & \text{ if } \vert x_n \vert > \sqrt{\frac{P_{\text{tx}}}{N}}, \end{cases}
\end{align}
where $n=1,\cdots,N$. Without loss of generality we assume that the maximal PA available power at each antenna is equal to $\sqrt{\frac{P_{\text{tx}}}{N}}$, where $P_{\text{tx}}$ denotes the total available power. This work can be easily extended to unequal available powers of the PAs among the antennas. 
A circuit-based characterization of the PA power consumption $P_{\text{PA}}$ from \cite{JeddaISWCS2015} provides the following expression
\begin{align}
P_{\text{PA}}= \sum_{n=1}^N \sqrt{\frac{P_{\text{tx}}}{N}} \E\left\lbrace \vert x_{\text{PA}_n}\vert\right\rbrace,
\end{align}
where $\sqrt{\frac{P_{\text{tx}}}{N}} $ and $ x_{\text{PA}_n}$ represent the PA saturation voltage and the output current at each antenna.

\section{Linear Relaxed Constant Envelope Precoder (L-RCE)}
\label{sec:linearprecoder}
This method is based on the transmit Wiener Filter approach \cite{Joham2005} with an additional term that hat penalizes instantaneous peak power that exceeds the PA saturation point. To minimize the PA distortions that affect the signals, the PA input signal has to be designed such that excursions of the designed transmit signal beyond the available peak power are discouraged. The optimization problem can be formulated as follows
\begin{align}
\min_{\matt P, f \in \mathtt{R}_+} \Psi(\matt P, f) &= \min_{\matt P, f \in \mathtt{R}_+} \left(1-\lambda \right) {\rm E} _{\vect s, \boldsymbol{\eta}} \left\lbrace \Vert f\left( \matt H \matt P \vect s + \boldsymbol{ \eta} \right)-\vect s \Vert^2_2 \right\rbrace \nonumber \\
&+ \lambda \tr\left( {\rm E}_{\vect s} \left\lbrace \left(\diag \left(\matt P \vect s \vect s^{\He} \matt P^{\He} \right) - \frac{P_{\text{tx}}}{N} \matt I_N \right)_+  \right\rbrace \right).
\label{eq:linear_opt_problem} 
\end{align}
The convex cost function $\Psi(\matt P, f)$ consists of a weighted sum of the mean-squared-error (MSE) between the that accounts for exceeding the maximal available PA power.
To better evaluate the penalty term we reformulate it as follows
\begin{align}
\tr &\left( {\rm E}_{\vect s} \left\lbrace \left(\diag \left(\matt P \vect s \vect s^{\He} \matt P^{\He} \right) - \frac{P_{\text{tx}}}{N} \matt I_N \right)_+  \right\rbrace \right)=\nonumber \\
& \tr\left( {\rm E}_{\vect s} \left\lbrace \Upsilon(\vect s) \matt P \vect s \vect s^{\He} \matt P^{\He}\Upsilon(\vect s) - \frac{P_{\text{tx}}}{N} \matt I_N   \right\rbrace \right),
\end{align}
where $\Upsilon(\vect s)$ is a diagonal selection matrix with 
\begin{align}
\Upsilon_{n,n}(\vect s)=\begin{cases} 0 & \text{if } \vect e^{\T}_n \matt P \vect s \vect s^{\He} \matt P^{\He} \vect e_n \leq \frac{P_{\text{tx}}}{N}, \\
1 & \text{otherwise.}
\end{cases}
\end{align}

The optimization problem in (\ref{eq:linear_opt_problem}) cannot be solved in a closed form. Thus, we resort to iterative methods such as the gradient descent method. To this end, the gradient expressions are derived.
The gradient with respect to $\matt P$ is expressed by
\begin{align}
\frac{\partial \Psi(\matt P, f)}{\partial \matt P} &= \left(1-\lambda \right)\left( f^2 \matt H^{\T} \matt H^* \matt P^* \matt C_s - f \matt H^{\T} \matt C_s \right) \nonumber \\
&+ \lambda  {\rm E}_{\vect s} \left\lbrace \Upsilon(\vect s) \matt P^* \vect s^* \vect s^{\T}\right\rbrace,
\label{eq:gradient_P_linear}
\end{align}
and the gradient with respect to $f$ is given by
\begin{align}
\frac{\partial \Psi(\matt P, f)}{\partial f} &= \left(1-\lambda \right) \Big( 2f \tr\left( \matt H \matt P \matt C_s \matt P^{\He} \matt H^{\He}+ \matt C_{\boldsymbol \eta} \right)\nonumber \\
&- 2 \tr\left( \Re\left\lbrace \matt H \matt P \matt C_s \right\rbrace\right) \Big).
\label{eq:derivative_f}
\end{align}
The optimal $f$ can be then calculated as 
\begin{align}
f&=\Bigg\vert\frac{\tr\left( \Re\left\lbrace \matt H \matt P \matt C_s \right\rbrace\right)}{\tr\left( \matt H \matt P \matt C_s \matt P^{\He} \matt H^{\He}+ \matt C_{\boldsymbol \eta} \right)}\Bigg\vert.
\end{align}
Note that the computation of (\ref{eq:gradient_P_linear}) is complex due to the second term. The mean value ${\rm E}_{\vect s}$ has to be computed for all possible input vectors $\vect s \in \mathcal{S}$, where the cardinality of $\mathcal{S}$ is given by $B^{M}$.  To reduce the algorithm complexity we resort to using a stochastic mean value.
Algorithm \ref{alg:linear_optproblem} describes the optimization steps to solve (\ref{eq:linear_opt_problem}).
\begin{algorithm}
\caption{L-RCE: Find optimal $\matt P$ and $f$}
\label{alg:linear_optproblem}
\begin{algorithmic}
\REQUIRE Channel matrix $\vect{H}$, $\mu$, $\epsilon$, $N_s$
\ENSURE $\check{\vect x}$
\STATE Generate $N_s$ random $\vect s$ vectors, $\vect s^{(1)}, \cdots, \vect s^{(N_s)}$,  from the set of all possible input vectors $\mathcal S$
	\STATE Compute $\Upsilon(\vect s^{(i)}), i=1,\cdots, N_s$
	\STATE $\matt P^{(0)} = \matt P_{\text{WF}}$ and $f^{(0)} = f_{\text{WF}}$
\WHILE{$\delta> \epsilon$} 
	\STATE $\matt P^{(n+1)} = \matt P^{(n)}-\mu \left( \frac{\partial \Psi(\matt P^{(n)}, f^{(n)})}{\partial \matt P}\right)^* $ \\
	\STATE $f^{(n+1)} = \Bigg\vert\frac{\tr\left( \Re\left\lbrace \matt H \matt P^{(n+1)} \matt C_s \right\rbrace\right)}{\tr\left( \matt H \matt P^{(n+1)} \matt C_s \matt P^{(n+1),\He} \matt H^{\He}+ \matt C_{\boldsymbol \eta} \right)}\Bigg\vert.$
	\STATE $\delta =\frac{ \|\Psi(\matt P^{(n+1)}, f^{(n+1)}) -\Psi(\matt P^{(n)}, f^{(n)}) \|} { \|\Psi(\matt P^{(n)}, f^{(n)})\|}$
	\IF{$\Psi(\matt P^{(n+1)}, f^{(n+1)}) >\Psi(\matt P^{(n)}, f^{(n)})$}
 	\STATE{$\mu = \mu/2$}  ~~~~~~// Step size adjustment
	\ELSE \STATE $n = n+1$
	\ENDIF
\ENDWHILE
\end{algorithmic}
\end{algorithm}

\section{Non-linear Relaxed Constant Envelope Precoder (NL-RCE)}
\label{sec:nonlinearprecoder}
Unlike the linear approach of the previous section, here we introduce a non-linear precoder in which for a given input vector $\vect s$ the following optimization problem has to be solved
\begin{align}
\min_{\vect x,f\in \mathtt{R}_+} \Phi\left( \vect x, f, \vect s \right) =& \min_{\vect x,f\in \mathtt{R}_+} {\rm E}_{\boldsymbol{\eta}} \left\lbrace \Vert f \left( \matt H \vect x +\boldsymbol{\eta}\right)  - \vect s\Vert^2_2\right\rbrace \nonumber  \\
&\text{s.t. } x_n \leq \sqrt{\frac{P_{\text{tx}}}{N}}, n=1,\cdots,N.
\label{eq:non-linear_opt_problem}
\end{align}
This symbol-wise precoding method is in general a non-linear mapping between each input vector $\vect s$ and its corresponding transmit vector $\vect x$.
The optimization problem in (\ref{eq:non-linear_opt_problem}) can be reformulated as follows
\begin{align}
&\min_{f\in \mathtt{R}_+} {\rm E}_{\check{\vect x }} \left\lbrace \min_{\vect x} \Vert f  \matt H \vect x  - \vect s\Vert^2_2\right\rbrace + f^2 \tr\left(\matt C_{\boldsymbol{\eta}} \right)  \nonumber  \\
&\text{s.t. } x_n \leq \sqrt{\frac{P_{\text{tx}}}{N}}, n=1,\cdots,N,
\end{align}
where $\check{\vect x } = \argmin_{\vect{x}} \Vert f  \matt H \vect x  - \vect s\Vert^2_2$. 
The transmit vector $\vect x$ has to be optimized for each given input vector $\vect s$, whereas the scalar $f$ has to be optimized jointly for all the input vectors $\vect s$. The scalar $f$ is applied at the receiver side where the input vector itself has to be estimated and therefore should not depend on the instantaneous value of $\vect s$.
For solving this optimization problem we resort again to the gradient descent method. Therefore, we derive the gradient expressions
\begin{align}
\frac{\partial \Phi\left( \vect x, f, \vect s \right)}{\partial \vect x} &= f^2 \matt H^{\T} \matt H^* \vect x^* - f \matt H^{\T} \vect s^{*},
\end{align}
and
\begin{align}
\frac{\partial \Phi\left( \vect x, f, \vect s \right)}{\partial  f} &=  {\rm E}_{\check{\vect x }}\left\lbrace  \frac{\partial \Vert f  \matt H \check{\vect x } - \vect s\Vert^2_2}{\partial f}\right\rbrace + 2f \tr\left(\matt C_{\boldsymbol{\eta}} \right)  \nonumber \\
&=  {\rm E}_{\check{\vect x }}\left\lbrace 2 f \tr\left( \matt H \check{\vect x } \check{\vect x }^{\He} \matt H^{\He} \right) - 2 \tr\left( \Re\left( \matt H \check{\vect x } \vect s^{\He}\right) \right) \right\rbrace\nonumber \\
&+2f \tr\left(\matt C_{\boldsymbol{\eta}} \right).
\end{align}
The optimal $f$ reads as
\begin{align}
f &= \Bigg\vert\frac{\E_{\check{\vect x }}\left\lbrace\tr\left( \Re\left( \matt H \check{\vect x } \vect s^{\He}\right) \right)\right\rbrace}{\E_{\check{\vect x }}\left\lbrace \tr\left( \matt H \check{\vect x } \check{\vect x }^{\He} \matt H^{\He} +\matt C_{\boldsymbol{\eta}} \right)\right\rbrace}\Bigg\vert.
\end{align}
The algorithm for solving $f$ is summarized in Algorithm \ref{alg:non-linear_optproblem_f}.

\begin{algorithm}
\caption{NL-RCE: Find optimal $f$}
\label{alg:non-linear_optproblem_f}
\begin{algorithmic}
\REQUIRE Channel matrix $\vect{H}$, $I$, $N_s$
\ENSURE Scaling factor $f$
\STATE $f = f_{\text{WF}}$
\FOR{$j:=1$ \TO $I$}
	\STATE Generate $N_s$ random $\vect s$ vectors, $\vect s^{(1)}, \cdots, \vect s^{(N_s)}$,  from the set of all possible input vectors $\mathcal S$
	\FOR{$i:=1$ \TO $N_s$}
	\STATE $\check{\vect x}^{(i)} = \argmin\limits_{\vect{x}} \Vert f  \matt H \vect x  - \vect s^{(i)}\Vert^2_2$
	\ENDFOR
	\STATE $f = \Bigg\vert\frac{\E_{\check{\vect x }}\left\lbrace\tr\left( \Re\left( \matt H \check{\vect x } \vect s^{\He}\right) \right)\right\rbrace}{\E_{\check{\vect x }}\left\lbrace \tr\left( \matt H \check{\vect x } \check{\vect x }^{\He} \matt H^{\He} +\matt C_{\boldsymbol{\eta}} \right)\right\rbrace}\Bigg\vert$
\ENDFOR
\end{algorithmic}
\end{algorithm}

Algorithm \ref{alg:non-linear_optproblem_x} describes the steps to find the optimal transmit vector $\check{\vect x}$ for a specific  $\vect s$. Note that this algorithm is also used in Algorithm \ref{alg:non-linear_optproblem_f} to find the optimal $f$. After optimizing the scaling factor $f$, the transmit vectors $\vect x$ are then optimized.

\begin{algorithm}
\caption{NL-RCE: Find optimal $\vect x$ for a given $\vect s$}
\label{alg:non-linear_optproblem_x}
\begin{algorithmic}
\REQUIRE Channel matrix $\vect{H}$, $\vect s$, $\mu$, $\epsilon$, $f$
\ENSURE $\check{\vect x}$
\STATE $\vect x^{(0)} = g_{\text{PA}}(\matt P_{\text{WF}} \vect s)$
\WHILE{$err > \epsilon$} 
	\STATE $\vect x^{(n+1)} = \vect x^{(n)}-\mu \left( \frac{\partial \Phi\left( \vect x, f, \vect s \right)}{\partial \vect x}\right)^* $ \\
	\STATE  $\vect x^{(n+1)} = g_{\text{PA}}\left(\vect x^{(n+1)} \right) $
	\STATE $err = \frac{\|\vect x^{(n+1)}-\vect x^{(n)}\|}{\|\vect x^{(n)}\|}$
	\IF{$\Phi\left( \vect x^{(n+1)}, f, \vect s \right) > \Phi\left( \vect x^{(n)}, f, \vect s \right)$}
 	\STATE{$\mu = \mu/2$} ~~~~~~// Step size adjustment
	\ELSE \STATE{$n = n+1$}
	\ENDIF
\ENDWHILE
\end{algorithmic}
\end{algorithm}

\section{Receive Processing}
\label{sec:receiver}
Once the designed precoder is applied, at the receive side, each user has to determine the appropriate scaling factor $f_m, m=1,\cdots,M$, independently.  We use the following blind estimation method for the scaling factor prior to decision ($f_m={\rm E}[|s_m|^2]/{\rm E}[s_mr_m^*]$) at the receiver that does not require any feedback or training from the base station and any knowledge of the noise plus interference power at the user terminal:
\begin{equation}
f_m= L \cdot \frac{{\rm E}\left[|{\rm Re}\{s\}|+|{\rm Im}\{s\}|\right]}{ \sum_{\ell=1}^L  |{\rm Re}\{r_m[\ell]\}|+|{\rm Im}\{r_m[\ell]\}|},
\end{equation}
with $L$ is  the length the received sequence. This formula is due to the fact
\begin{equation}
\begin{aligned}
&|{\rm Re/Im}\{r_m\}|=|{\rm Re/Im}\{f_m^{-1} s_m+ e_m\}|  \\
& \stackrel{\textrm{w/ high prob. at SINR $\gg$ 1}}{=} |{\rm Re/Im}\{f_m^{-1} s_m\}| +  {\rm Re/Im}\{e_m\},
\end{aligned}
\end{equation}
meaning that with zero-mean noise plus interference $e_m$ we have ${\rm E}[|{\rm Re/Im}\{r_m\}|] \approx f_m^{-1} {\rm E}[|{\rm Re/Im}\{s_m\}|] $ .

\section{Simulation Results}
\label{sec:results}
\begin{figure}[htb]
  \centering
  \centerline{\includegraphics[width=1\columnwidth]{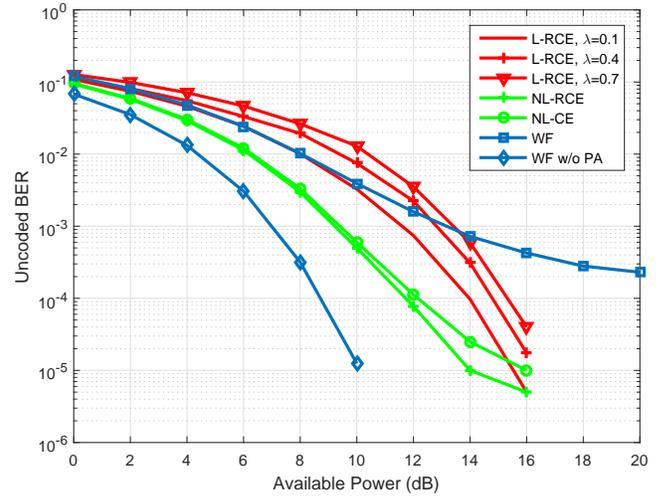}}
  \caption{BER vs. Available Power for different precoder designs.}
    \label{avai_P}
\end{figure}

\begin{figure}[htb]
  \centering
  \centerline{\includegraphics[width=1\columnwidth]{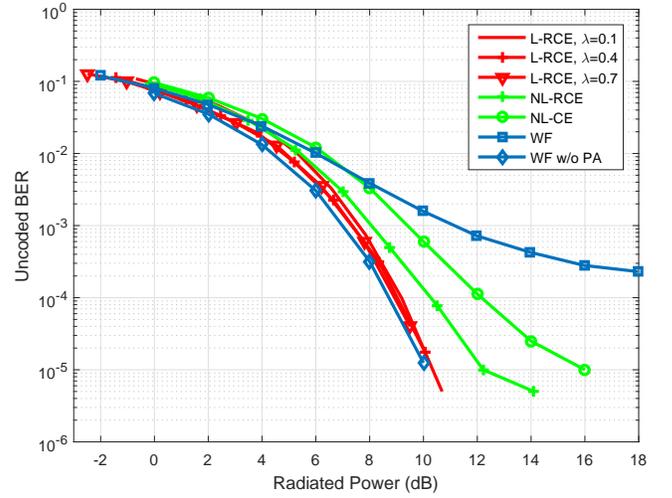}}
    \caption{BER vs. Radiated Power for different precoder designs.}  
        \label{rad_P}
\end{figure}

\begin{figure}[htb]
  \centering
  \centerline{\includegraphics[width=1\columnwidth]{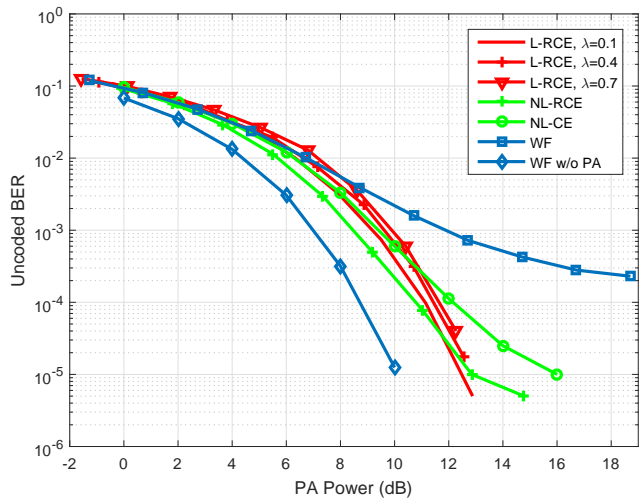}}

    \caption{BER vs. PA Power for different precoder designs.}    
        \label{PA_P}  
\end{figure}

For the simulations, we assume a
BS with $N = 100$ antennas serving $M = 10$ single-antenna
users. The channel $\vect H$ is composed of i.i.d. Gaussian
random variables with zero-mean and unit variance. 
All the simulation results are obtained with $N_b=10^3$ transmit symbols per channel use. 
The additive
noise is also i.i.d with variance one at each antenna. The assumed modulation scheme is 16QAM. 
The performance metrics are on the one hand the uncoded bit-error-ratio (BER) and on the other hand the following characteristic powers that we aim to minimize:
\begin{itemize}
\item available power $P_{\text{tx}}$ in Fig.~\ref{avai_P}: the maximum power that can be delivered by the amplifier and reflects the complexity and the cost of the PA 
\item radiated power $P_{\text{r}}$ in Fig.~\ref{rad_P}: the effective radio frequency (RF) power transmitted (${\rm E}[\left\| \vect x_{\rm PA} \right\|^2_2]$) that might act as interference for other systems operating in the same frequency band
\item PA consumed power $P_{\text{PA}}$ in Fig.~\ref{PA_P}: the total power consumption of the amplifier and characterizes the overall power efficiency. 
\end{itemize}
The curve labeled L-RCE (linear relaxed CE) refers to the proposed linear precoder design with different weighting coefficients $\lambda$ as opposed to the standard Wiener filter (WF) design from \cite{Joham2005}, the non-linear CE (NL-CE) precoding proposed in \cite{Larsson2013}, and its relaxed version (NL-RCE) discussed in this paper. In addition, the ideal performance without PA distortion (WF, w/o PA) is plotted for comparison. 

Interestingly, taking $\lambda=0.1$ in the proposed approach provides nearly the best results among linear methods in terms of all characteristic powers, especially when compared to the WF approach. Additionally, we observe from Fig.~\ref{rad_P} that the degradation of L-RCE compared to the ideal case (WF, w/o PA) is negligible. When comparing the symbol-wise non-linear methods, i.e., the NL-CE method versus the relaxed version NL-RCE, we observe an improvement in terms of radiated power and also PA power consumption, suggesting the utilization of the linear region of the PA. Additionally, while the non-linear precoding methods (NL-CE and NL-RCE) have some advantage in terms of available power compared to the linear method L-RCE presented here, they require higher radiated power, and thus potentially higher interference to other systems and significantly higher computational complexity. From a practical point of view, we conclude that linear precoding taking into account the effects of PA clipping might perform sufficiently well.

\section{Conclusion}
\label{sec:conclusion}

This paper presented linear and non-linear precoding methods for improving the performance of a downlink multi-user MISO system when the signal is affected by non-linear amplifiers at
the BS transmitter. Using an appropriate PA model, simulations
showed that relaxing the power constraint of existing CE precoding to be an instantaneous peak power constraint instead of the constant power constraint is advantageous in terms of radiated power and overall-power consumption. Furthermore, an appropriate linear precoder design taking the clipping effect into account is proposed, and is shown to have comparable performance in terms of power efficiency to the existing CE approach with much lower processing complexity.



%

\bibliographystyle{IEEEtran}
\bibliography{IEEEabrv,refs}

\end{document}